\documentclass[aps,prl,reprint,preprintnumbers,showpacs,showkeys,superscriptaddress,floatfix,nofootinbib]{revtex4-1}
\usepackage{epsfig}
\usepackage{amsfonts}
\usepackage{latexsym}
\usepackage{amsmath,amssymb}
\usepackage{mathrsfs}
\usepackage{hyperref}
\usepackage{setspace}
\usepackage{color}
\usepackage{bm}
\usepackage{slashed}
\usepackage{braket}
\newcommand{\tf}{t_{\rm f}}

\begin{document}

\unitlength = 1mm
\begin{flushright}
KOBE-COSMO-21-06
\end{flushright}


\title{Indirect detection of gravitons through quantum entanglement}

\author{Sugumi Kanno}
\affiliation{Department of Physics, Kyushu University, 744 Motooka, Nishi-ku, Fukuoka 819-0395, Japan}
\author{Jiro Soda}
\affiliation{Department of Physics, Kobe University, Kobe 657-8501, Japan}
\author{Junsei Tokuda}
\affiliation{Department of Physics, Kobe University, Kobe 657-8501, Japan}


\begin{abstract}
We propose an experiment that the entanglement between two macroscopic mirrors suspended at the end of an equal-arm interferometer is destroyed by the noise of gravitons through bremsstrahlung. By calculating the correlation function of the noise, we obtain the decoherence time from the decoherence functional. We estimate that the decoherence time induced by the noise of gravitons in squeezed states stemming from inflation is approximately 20 seconds for 40 km long arms and 40 kg mirrors. Our analysis shows that observation of the decoherence time of quantum entanglement has the potential to detect gravitons indirectly. 
This indirect detection of gravitons would give strong evidence of quantum gravity.
\end{abstract}




\maketitle

\noindent
{\it Introduction.---}
It is widely believed that gravity should be quantized because other forces are quantized. Nevertheless, there is no satisfactory quantum theory of gravity. Some physicists think that gravity may not be quantized after all~\cite{Jacobson:1995ab}. Thus, to find an experimental evidence of quantum gravity is quite important. A clear consequence of quantum theory of gravity would be the existence of gravitons. Hence, it is desired to come up with a novel way to observe gravitons experimentally. 

It has been known that gravitons are imperceptible. In fact, Dyson has conjectured that no conceivable experiment in our universe can detect a single graviton~\cite{DYSON:2013jra,Rothman:2006fp}. 
If it is true, we need to seek alternative ways for confirming the existence of gravitons.
One possible way is to focus on an inflationary scenario. It is believed that primordial gravitational waves can be generated during inflation from quantum fluctuations of geometry. If we succeed in observing the primordial gravitational waves, it would imply a discovery of gravitons~\cite{Kanno:2017dci,Kanno:2018cuk,Kanno:2019gqw}. However, even if the primordial gravitational waves arrive at the interferometers, the direct detection of the gravitons is difficult with current experimental techniques. Another possible way is to focus on the statistical property of primordial gravitational waves.  The state of the primordial gravitational waves becomes squeezed during inflation~\cite{Grishchuk:1974ny, Grishchuk:1990bj}. Unfortunately, it is pointed out that observing the squeezed states is also practically impossible~ \cite{Allen:1999xw}.

Let us recall the history of discovery of atoms. Einstein used Brownian motion to deduce the existence of atoms.
In the same way, instead of direct detection of gravitons, indirect search for gravitons might be possible.  
Recently, the noise induced by gravitons is discussed
 in \cite{Parikh:2020nrd,Parikh:2020kfh,Parikh:2020fhy,Haba:2020jqs}.
An indirect detection of gravitons by making use of the process of decoherence through the noise of gravitons is also proposed in \cite{Riedel:2013yca, Suzuki:2015nva, Kanno:2020usf}. However, no feasible experiment has been proposed yet.
The goal of this letter is to propose a feasible experimental setup for indirect detection of gravitons in the squeezed states stemming from inflation. In the following, we work in the natural unit: $c=\hbar =1$.

\begin{figure}[ht]
\vspace{-0.4cm}
\includegraphics[width=7cm]{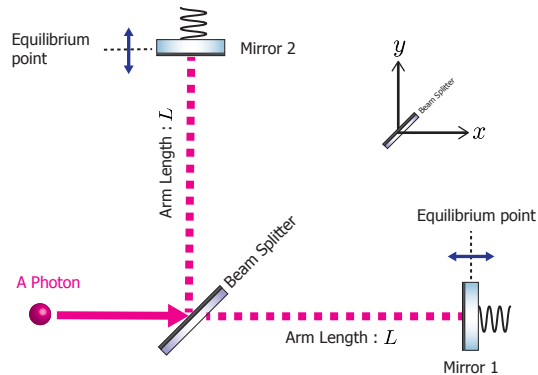}
\centering
\vspace{-0.3cm}
\caption{The proposed setup: an equal-arm Michelson interferometer for a single photon where there is a macroscopic suspended mirror at the end of each arm} 
\label{fig1}
\end{figure}
\noindent
{\it Experimental setups.---} The proposed setup, shown in FIG.~\ref{fig1} consists of an equal-arm Michelson interferometer which has a macroscopic suspended mirror at the end of each arm.  An incident photon (photon beams with very low intensity) is injected from left and the beam splitter converts the photon into a superposition state of being in both the upper and lower arms simultaneously until it is detected by the oscillation of either mirror. Here we assume the oscillation of the upper and lower mirrors ($2$ and $1$) as a semiclassical state described by
\begin{eqnarray}
&&\vec\xi_1 (t)  = (\xi_1,0,0)\,,\quad \xi_1=  A \cos \omega t \,,\nonumber\\
&& \vec\xi_2 (t) = (0,\xi_2,0)\,,\quad \xi_2=  A \cos \omega t \,,
\end{eqnarray}
where the amplitude of the oscillation $A$ and the angular frequency $\omega$ of both mirrors are set to the same value for simplicity. Let $H_1$ and $H_2$ denote each Hilbert space of mirrors. Then the Hilbert space for the combined system of the two mirrors is $H_1\otimes H_2$. Let $H_1$ be spanned by a basis $\{ |0\rangle , \big| \vec\xi_1  \rangle\}$ and $H_2$ be spanned by $\{ |0\rangle , \big| \vec\xi_2  \rangle\}$ where $|0\rangle$ is the vacuum state in equilibrium and $|\vec\xi_{\rm i}\rangle$ (${\rm i}=1,2$) represents a coherent state as an excited state induced by a single photon.
 After an incident photon is injected, the state of both mirrors with mass $m$ is described by the maximally entangled state where mirror 1 oscillates and mirror 2 is in equilibrium or mirror 1 in equilibrium and mirror 2 oscillates 
\begin{equation}
      \frac{1}{\sqrt{2}}\, \big| \vec\xi_1  \rangle  \otimes \big| 0 \rangle  +  \frac{1}{\sqrt{2}}\,  \big| 0  \rangle \otimes \big| \vec\xi_2 \rangle\,.
\end{equation}

Now, let us consider the influence of environmental quantum gravitational fields (gravitons) on the system of the mirrors. We assume that the initial total system is given by
$
       \big| 0  \rangle \otimes \big| 0 \rangle \otimes \big| {\rm vac} \rangle\,,
$
where $|{\rm vac }\rangle$ represents the vacuum state of gravitons. After the incident photon is injected, the system of the mirrors gets entangled and the total system becomes
\begin{equation}
      \big| \psi (t_{\rm i})  \rangle  = \left\{ \frac{1}{\sqrt{2}} \big| \vec\xi_1   \rangle 
       \otimes \big| 0  \rangle
  +  \frac{1}{\sqrt{2}}  \big| 0   \rangle \otimes \big| \vec\xi_2  \rangle 
  \right\} \otimes \big| {\rm vac}  \rangle\,.
\end{equation} 
where $t_{\rm i}$ is the initial time. If we focus on the system of the mirrors, the reduced density operator is obtained by tracing out the degree of freedom of environmental gravitons such as
\begin{eqnarray}
  \rho_m (t_{\rm i})
 &=& {\rm Tr}_{\rm grav}|\psi (t_{\rm i}) \rangle\langle\psi (t_{\rm i}) |     \nonumber\\
  &=&    \rho_{11} (t_{\rm i} )   +  \rho_{22} (t_{\rm i} )  + \rho_{12} (t_{\rm i} )  + \rho_{21} (t_{\rm i} )       \,,\qquad
\end{eqnarray}
where we defined $\rho_{11}\equiv |\vec\xi_1\rangle|0\rangle\langle\vec\xi_1|\langle 0|$ etc. The interference term $\rho_{12}+ \rho_{21}$ shows the initial entangled state between the mirrors. \\\indent
The total system evolves in time according to the Langevin equation of geodesic deviation of the mirrors in the presence of gravitons~\cite{Parikh:2020nrd, Kanno:2020usf}. Since each mirror interacts with the surrounding gravitons as we will see in Eq.~(\ref{action}), the vacuum state of the gravitons changes due to bremsstralung in accordance with the oscillation of the mirror on either side. The total system is then  found to be
\begin{eqnarray}
      \big| \psi (t_{\rm f})\rangle &=& \frac{1}{\sqrt{2}} \big| \vec\xi_1  \rangle  \otimes \big| 0 \rangle\otimes \big| {\rm vac}; {\vec\xi_1} \rangle \nonumber\\
&&+  \frac{1}{\sqrt{2}}  \big| 0  \rangle \otimes \big| \vec\xi_2 \rangle  \otimes \big| {\rm vac} ;{\vec\xi_2}\rangle\,,
\end{eqnarray}
where $t_{\rm f}$ is the final time and $|\rm{vac};\vec\xi_i\rangle$ represents the state of gravitons after the interaction with the oscillating mirror ${\rm i}$. We ignored the force of radiation reaction of the mirror because the gravitational back reaction is negligible and then the state of each mirror remains unchanged. The point here is that the system of the mirrors gets entangled with environmental gravitons due to the matter-gravity interaction. In other words, the decoherence of the system occurs due to bremsstralung of gravitons. The effect of the bremsstralung of gravitons is expressed by the reduced density operator of the form
\begin{eqnarray}
  \rho_m (t_f ) 
  &=&    \rho_{11} (t_{\rm i} )   +  \rho_{22} (t_{\rm i} )  \nonumber\\
 && + \exp(i\Phi) \rho_{12} (t_{\rm i} )  + \exp(-i\Phi^*)\rho_{21} (t_{\rm i} )       \,,\qquad
\end{eqnarray}
where the influence of the gravitons on the reduced system is expressed by the influence functional $\Phi$. The imaginary part of the influential functional suppress the interference term and that is referred  to as the decoherence functional $\Gamma = {\rm Im}\Phi$~\cite{Feynman:1963fq,Breuer}. If $\Gamma$ is calculated, we can read off the decoherence time from it. In order to quantify the change of entanglement over time, the entanglement negativity as a measure of entanglement is computed as
\begin{eqnarray}
  {\cal N}
  =    \big|\rho_{12} (t_{\rm i} ) \big|    \exp(-\Gamma)   \  .
\end{eqnarray}
By observing the decoherence time or the change of entanglement over time, we could detect gravitons indirectly. In the next section, we see how the mirrors interact with gravitons.\\

\noindent
{\it Action for the total system.---}
Let us consider the action for gravitational waves in the Minkowski space first. The metric describing gravitational waves in the transverse traceless gauge is expressed as\begin{eqnarray}
ds^2=-dt^2+(\delta_{ij}+h_{ij}) dx^idx^j\,\,,
\label{metric}
\end{eqnarray}
where $t$ is the time, $x_i$ are spatial coordinates, $\delta_{ij}$ and $h_{ij}$ are the Kronecker delta 
and the metric perturbations which satisfy the transverse traceless conditions $h_{ij,\,j}=h_{ii}=0$. The indices $(i,j)$ run from $1$ to $3$.  
Substituting the metric Eq.~(\ref{metric}) into the Einstein-Hilbert action, we obtain the quadratic action for the metric perturbations
\begin{eqnarray}
S_g=  \frac{M_{\rm p}^2}{8}\int{\rm d}^4x\,
\left[\,\dot{h}^{ij} \dot{h}_{ij}   -   h^{ij,k}\,     h_{ij,k}
\,\right]  \label{G_action}\, ,
\end{eqnarray}
where the reduced Planck mass is defined by $M_{\rm p}^{-2}=8\pi G$ and a dot denotes the derivative with respect to the time.
We can expand the metric field $h_{ij}( x^i, t)$ in terms of the Fourier modes
\begin{eqnarray}
h_{ij}(x^i ,t ) = \frac{2 }{M_{\rm p}\sqrt{V}}\sum_{{\bf k}, B} \ h^B_{\bf{k}}(t)\,e^{i {\bf k} \cdot {\bf x}} \ e_{ij}^B(\bf k)  \,,\qquad
\label{fourier}
\end{eqnarray}
where we introduced the polarization tensor $e^B_{ij}({\bf k})$ normalized as $e^{*B}_{ij}({\bf k}) e^C_{ij} ({\bf k})= \delta^{BC}$. 
Here, the index $B$ denotes the linear polarization modes $B=+,\times$. 
Note that  we consider finite volume $V=L_{x }L_{y}L_{z}$ and 
 discretize the ${\bf k}$-mode with a width ${\bf k} = \left(2\pi  n_x/L_x\,,2\pi  n_y/L_y\,,2\pi  n_z/L_z\right)$ where ${\bf n}=(n_x , n_y , n_z)$ 
are integers.  
Note that we used $k=|\bf k|$. 
We then see that a gravitational wave consists of an infinite number of harmonic oscillators. \\\indent
Next, we consider the action for two mirrors. Here, we regard the mirror as a point particle effectively because the dynamical degree of the center of mass is essential for discussing the noise of gravitons later. Then we consider how the particle feels gravitational waves. In fact, a single particle does not feel the gravitational waves because of the Einstein's equivalence principle at least classically. Thus, we need to consider geodesic deviation between the two particles in order to see the effect of gravity on them. We introduce the Fermi normal coordinates for calculating the geodesic deviation between the time-like geodesics of two particles $\gamma_{\tau}$ and $\gamma_{\tau'}$. The Fermi normal coordinates are local inertial coordinates that are adapted to a geodesic. We expand the coordinate along the time-like and space-like geodesics that are orthogonal to each other at the position of the beam splitter as shown in FIG.~\ref{fig1} where the space-like geodesics are simplified as $(x,y)$-plane.
The dynamics of the $\gamma_{\tau}$ and $\gamma_{\tau'}$ is described by the position $\xi_{\rm i}(t) , {\rm i}=1,2$ respectively in the vicinity of the origin of the beam splitter.
The action for a particle along $\gamma_\tau$ is given by
\begin{eqnarray}
\hspace{-5mm}
  S_p =  - m\int_{\gamma_{\tau}}d\tau = -m\int_{\gamma_{\tau}} dt\sqrt{-g_{\mu\nu}(t,\xi^i(t))\dot{\xi}^\mu\dot{\xi}^\nu}    \,,
 \label{action1}
\end{eqnarray}
where $\tau$ is proper time and the position of the particle is represented by $\xi^\mu=(t,\xi^i(t))$. 
 The metric $g_{\mu\nu}$ up to the second order of arbitrary position $x^i$ in the Fermi normal coordinates is computed as
\begin{eqnarray}
ds^2 &\simeq& \left(- 1- R_{0i0j}\,x^i x^j \right) dt^2 -\frac{4}{3} R_{0jik}\,x^j x^k\, dt dx^i \nonumber\\
&&+\left(\delta_{ij} -\frac{1}{3}  R_{ikj\ell}\,x^k x^\ell \right) dx^i dx^j \ .
\label{metric2}
\end{eqnarray}
Here the Riemann tensor is evaluated at the origin $x^i =0$ in the Fermi normal coordinate system. 
Because the Riemann tensor $R_{0i0j}$ is gauge invariant at the leading order in the metric fluctuation $h_{ij}$, we can evaluate it in the transverse traceless gauge and obtain $R_{0i0j}(0,t)=- \ddot{h}_{ij} (0,t)/2$.
Substituting the metric~(\ref{metric2}) into the action~(\ref{action1}),  we can read off the interaction between the particles and gravitational waves as
\begin{eqnarray}
S_{\rm int} &\simeq& \int dt \left[
\frac{m}{4} \ddot{h}_{11} (0,t)\,{\xi^1_1}^2  +  \frac{m}{4} \ddot{h}_{22} (0,t)\,{\xi^2_2}^2 \right] \,.
\label{action}
\end{eqnarray}
where $x,y$ components are expressed by $1,2$ respectively. We find that a qubic derivative interaction between gravitational waves and geodesic deviation appeared 
in the action. We can then derive the Langevin equation of geodesic deviation of the mirrors in the presence of gravitons from this action~\cite{Parikh:2020nrd, Kanno:2020usf}.
Note that the geodesic deviation of particles in the graviton background is studied in~\cite{DeLorenci:2014vwa, Quinones:2017wka}.  Note also that a particle motion in the graviton background is discussed in \cite{Oniga:2015lro,Oniga:2017pyq} where a different form of interaction is used.\\

\noindent
{\it Quantization and the noise induced by gravitons.---}
Now we canonically quantize the total system. For simplicity, we work in the interaction picture below. We promote the free metric field $h^A({\bf k},t)$ to the operator $\hat h^A({\bf k},t)$ in terms of the creation and annihilation operators such as
\begin{align}
\hat h^A({\bf k},t)
      =\hat a_A({\bf k})u_k(t)+\hat a^\dag_A(-{\bf k})u^*_k(t) \   \,,
\label{hq}
\end{align} 
where the creation and annihilation operators satisfy the standard commutation relations
$
[\,{\hat a}_A({\bf k})\,,\,\hat{a}^\dag_{A'}({\bf k'})\,]=\delta_{{\bf k},{\bf{k}'}}\delta_{AA'}  \,,         
$
and $u_k(t)$ denotes a mode function properly normalized as
$
\dot u_k(t)u_k^*(t)-u_k(t)\dot u_k^*(t)=-i\,.
$
The Minkowski vacuum $\ket{0}$ is defined by $\hat a_A({\bf k}) \ket{0} =0$, with choosing the mode function as 
$
u_k(t)=\frac{1}{\sqrt{2k}}e^{-ikt}\equiv u_k^{\rm M}(t)   \ .
$
From the Langevin equation of geodesic deviation of the mirrors in the presence of gravitons, the noise of gravitons is identified as~\cite{Kanno:2020usf}
\begin{eqnarray}
\hat{N}_{ij}(t)\equiv\frac{1}{M_{\rm p} \sqrt{V}} \sum_A \sum_{{\bf k}\leq\Omega_{\rm m}} k^2  e^{A}_{ij}({\bf k} ) \hat h^A({\bf k},t)\,,
\qquad
\label{noise}
\end{eqnarray}
where $\sum_{{\bf k}\leq\Omega_{\rm m}}$  represents the mode sum with the UV cutoff. This noise of gravitons always exists if the gravitational waves are quantized. \\

\begin{figure}[ht]
\vspace{-0.3cm}
\includegraphics[width=8.5cm]{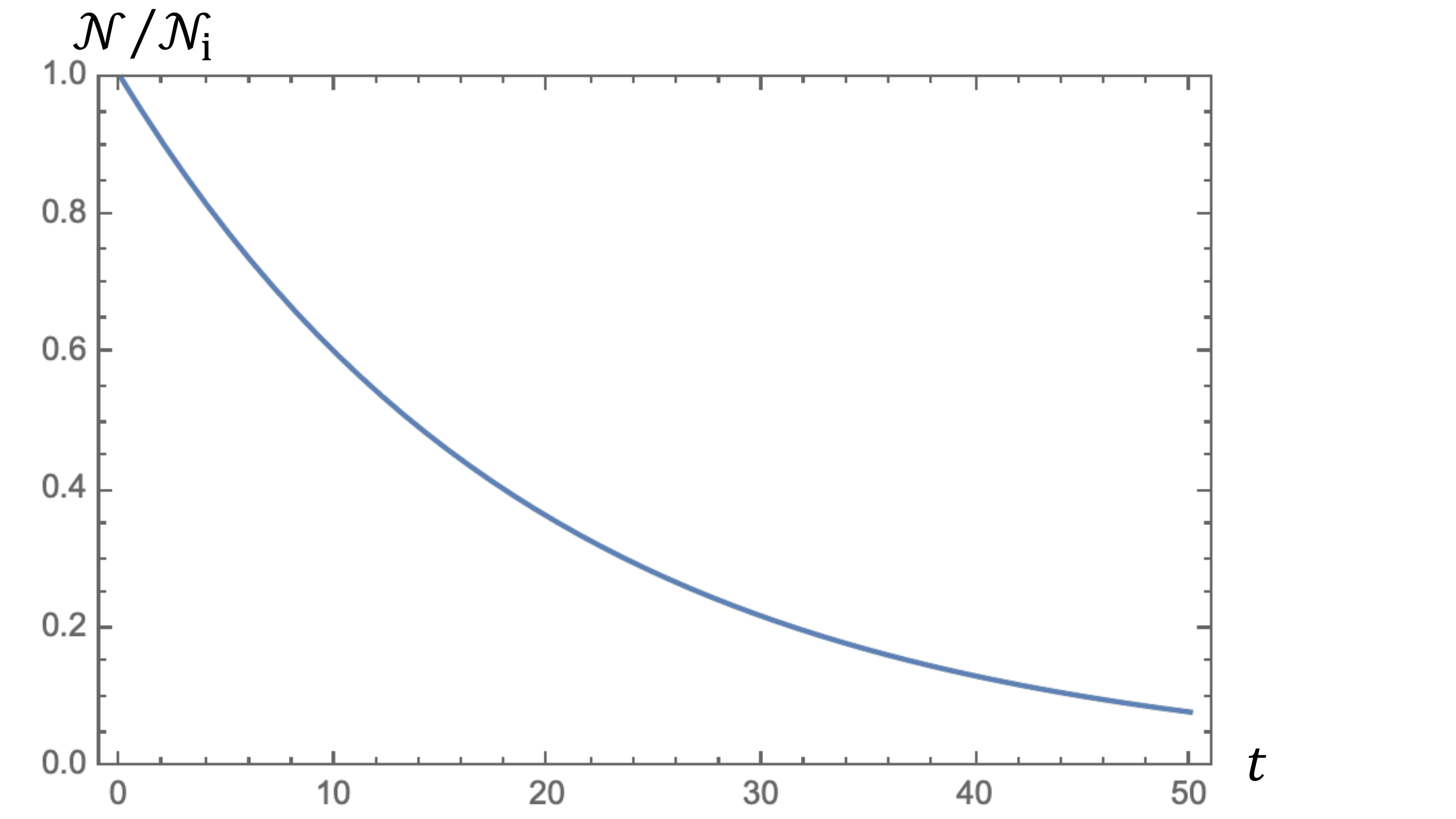}
\centering
\vspace{-0.3cm}
\caption{Time evolution of negativity normalized by the initial value ${\cal N}_{\rm i}$ for $\omega = 1$ kHz, $L=40$ km, $m=40$ kg, $f_{\rm c} =10^9$ Hz. The negativity decays with the decoherence time $20$ s.} 
\label{fig2}
\end{figure}

\noindent
{\it The decoherence functional and the decoherence time.---}
Since we found that gravitons are quantum fluctuations of gravitational waves in the form of the noise  Eq.~(\ref{noise}), we calculate the decoherence functional and then find the necessary experimental setups for obtaining a measurable time of decoherence due to the noise of gravitons. For this purpose, we consider squeezed states $|\zeta\rangle$ below. For instance, the state of the primordial gravitational waves becomes squeezed during inflation~\cite{Grishchuk:1974ny, Grishchuk:1990bj}. The decoherence functional  reads~\cite{Breuer, Kanno:2020usf}
\begin{eqnarray}
\Gamma (\tf) &\approx&\frac{m^2}{8}\int^{\tf}_0\mathrm{d}t\,\Delta(\xi^i\xi^j)(t)\int^{\tf}_0\mathrm{d}t'\,\Delta(\xi^k\xi^\ell)(t')   \nonumber\\
&&\times \Big\langle\left\{\hat N_{ij}(t),\,\hat N_{k\ell}(t')\right\}\Big\rangle\,.
\label{rate}
\end{eqnarray}
Here $\Delta(\xi^i\xi^j)(t)=\xi^i_1(t)\xi^j_1(t)-\xi^i_2(t)\xi^j_2(t)$ denotes a difference of $\xi^i(t)\xi^j(t)$ in the superposition. 
The anticommutator correlation function of $N_{ij}(t)$ in the squeezed state can be computed 
 in the infinite volume limit $L_x,L_y,L_z\to\infty$ as
\begin{eqnarray}
&& \Big\langle \zeta\Big|\left\{\hat N_{ij}(t),\,\hat N_{k\ell}(t')\right\}\Big|\zeta\Big\rangle 
\nonumber \\
&&  = \left( \delta_{ik} \delta_{j\ell} +  \delta_{i\ell} \delta_{jk} 
                 - \frac{2}{3}  \delta_{ij}  \delta_{k\ell}      \right)  \frac{F(t-t')}{10\pi^2M_{\rm p}^2}   \,, \quad
\label{eq:noiseamp3}
\end{eqnarray}
where we defined the anticommutator symbol $\{\cdot\,,\cdot\}$ as $\{\hat X,\hat Y\}\equiv(\hat X\hat Y+\hat Y\hat X)/2$ and 
\begin{eqnarray}
F(t-t')\equiv \int^{\Omega_m}_0\mathrm{d}k\,k^6\mathrm{Re}\left[u^{\rm sq}_k(t){u^{\rm sq}_k}^*(t')\right] \ .
\end{eqnarray}
Here, the mode function in the squeezed state is given by
\begin{eqnarray}
u^{\rm sq}_k(t)\equiv u^{\rm M}_k(t)\cosh r_k-e^{-i\varphi_k}{u_k^{\rm M}}^*(t)\sinh r_k\,.
\end{eqnarray}
In general, the squeezing parameter $r_k$ and the phase $\varphi_k$ depend on $k$ but 
we assume $\varphi_k =\varphi$ for simplicity and $r_k \gg 1$ which corresponds to the end of inflation. In the conventional inflationary scenario, we obtain $\sinh 2r_k \simeq \cosh 2r_k \simeq \left( k_{\rm c}/k \right)^4$ where $k_{\rm c}=2\pi f_{\rm c}$ and  $f_{\rm c}$ is the cutoff frequency of primordial gravitational waves. The bounds on the cutoff frequency from the CMB is $f_{\rm c}\lesssim 10^9$ Hz~\cite{Maggiore:1999vm}. For alternatives to inflation, the bounds become $f_{\rm c}\lesssim 4.3\times 10^{10}$ Hz~\cite{Maggiore:1999vm}.
We have a maximum value of the noise correlation (\ref{eq:noiseamp3}) for $\varphi=\pi$. 
By a change of variables $y= k /\Omega_{\rm m}, x= \Omega_{\rm m} (t-t')$, we can evaluate the integral as
\begin{eqnarray}
F(x) = k_{\rm c}^4 \Omega_m^2 \frac{x\sin x +\cos x -1  }{x^2}
\ .\label{eq:squezamp}
\end{eqnarray}
In our setup, the decoherence functional (\ref{rate}) is
\begin{eqnarray}
&&\Gamma = \frac{m^2 }{120\pi^2 M_{\rm p}^2} \int_0^{\tf} dt \Delta\xi_1^2 (t) \int_0^{\tf}  dt'\Delta\xi_1^2 (t') F( t-t' )  \nonumber\\
     && +    \frac{m^2 }{120\pi^2 M_{\rm p}^2} \int_0^{\tf} dt \Delta\xi_2^2 (t) \int_0^{\tf} dt' \Delta\xi_2^2 (t') F( t-t' ) \nonumber\\
  &&             -  \frac{m^2  }{120\pi^2 M_{\rm p}^2} \int_0^{\tf} dt \Delta\xi_1^2 (t) \int_0^{\tf} dt' \Delta\xi_2^2 (t') F( t-t' )\,,
\label{double}
\end{eqnarray}
where $ \Delta\xi_1^2,  \Delta\xi_1^2$ are
\begin{eqnarray}
    \Delta\xi_1^2 (t)  &=&  \left( L + A \cos \omega t \right)^2  -  L^2\,,     \\
       \Delta\xi_2^2 (t)  &=&  L^2  -  \left( L + A \cos \omega t \right)^2\,.     
\end{eqnarray}
Here, we stress that the decoherece process we considered is non-Markovian, that is, we took into account the correlated noise of gravitons. In the literature~\cite{Blencowe:2012mp,Anastopoulos:2013zya}, however, the decoherence rate is computed in the Markovian approximation.\\\indent
Let us evaluate the double integrals in Eq.~(\ref{double}). Since the amplitude of oscillation induced by a photon is negligible compared with the arm length $A\ll L$, $\Delta\xi_i^2\sim 2LA\cos\omega t$ and we obtain
 \begin{eqnarray}
\Gamma = \frac{m^2  }{10\pi^2 M_{\rm p}^2}  \left( A L \right)^2
I  \  , 
\end{eqnarray}
where 
\begin{eqnarray}
I&=&  \ \frac{1}{ \Omega_{\rm m}^2} \int_0^{\tf} dx \cos \beta x \nonumber\\
&&  \times \int_0^{\tf}dx' \cos \beta x' F( x - x')  \,.
\end{eqnarray}
Here we defined the dimensionless time and frequency $ x= \Omega_{\rm m} t$, $\beta=\omega/\Omega_{\rm m}$, respectively.
After the integration, we find the following terms become dominant for large $x$ 
\begin{eqnarray}
\frac{I(x)}{k_c^4} 
&\simeq& -\frac{1}{2}\beta x\,{\rm Si} \left( (1+\beta) x\right)  \nonumber\\
&&
+ \frac{1}{2}\beta x\,{\rm Si}\left( (1-\beta) x\right) +\beta x\,\rm{Si}(\beta x)\,,
\end{eqnarray}
where the sine integral is defined by
$
{\rm Si} (x) = \int^x_0 \frac{\sin t}{t} dt
$
whose value at $x=\infty$ is $\pi /2$. 
Thus, the decoherence functional is found to be
\begin{eqnarray}
\Gamma = \frac{4\pi^3}{5}\left(\frac{m}{M_{\rm p}}\right)^2 \left(  L f_{\rm c} \right)^4 \left( \frac{A}{L} \right)^2         N     \  , 
\end{eqnarray}
where we defined $N=\omega t_f $. 
Remarkably, the resultant decoherence functional is independent of the UV cutoff $\Omega_{\rm m}$, which indicates the result is reliable.
If we consider conventional inflation and take the parameters, $\omega = 1$ kHz, $L=40$ km, $m=40$ kg, $f_{\rm c} =10^9$ Hz, $N=2\times10^4$,
we obtain $\Gamma \simeq 1$ where the amplitude of oscillation is supposed to be about 10 times zero-point fluctuations $A=10/\sqrt{2m\omega}$~\cite{Gely:2021fhv}. Then the decoherence time ${\tf}$ is approximately $20$ s. In this case, the time evolution of negativity is plotted in FIG.~\ref{fig2}.
 If we increase the arm length $L$ by $10^2$, the decoherence time becomes $2$ ms.\\
\indent
For alternatives to inflation ($f_{\rm c}=4.3\times 10^{10}$ Hz~\cite{Maggiore:1999vm}), the decoherence time becomes $6$ $\mu$s with the same parameters.\\

\noindent
{\it Conclusion.---}
We performed a detailed study of the experimental setup for detecting gravitons indirectly by observing the decoherence time of the entanglement between two macroscopic mirrors suspended at the end of an equal-arm  interferometer. In principle the proposed setup can be realized in near future. We found that longer arms or heavier mirrors make the decoherence time shorter. We estimated that the decoherence time induced by the noise of gravitons in the squeezed state stemming from inflation is approximately 20 seconds for 40 km long arms and 40 kg mirrors. The arm length is 1 order of magnitude longer than that of LIGO. To generate entanglement between 40 kilogram mirrors appears to be within reach of current technology~\cite{Yu:2020ece}.
The other values used in the estimation is achieved in laboratory to date. There remains an issue to suppress other sources of the decoherence.  While very demanding, the indirect detection of gravitons  appears to be within reach in near future.

\noindent
{\it Acknowledgments.---}
We would like to thank Nobuyuki Matsumoto for useful discussion and suggestions.
S.\,K. was supported by the Japan Society for the Promotion of Science (JSPS) KAKENHI Grant Number JP18H05862.
J.\,S. was in part supported by JSPS KAKENHI Grant Numbers JP17H02894, JP17K18778, JP20H01902.
J.\,T. was supported by  JSPS Postdoctoral Fellowship No. 202000912.


\begin{thebibliography}{99}

\bibitem{Jacobson:1995ab}
T.~Jacobson,
Phys. Rev. Lett. \textbf{75}, 1260-1263 (1995)
[arXiv:gr-qc/9504004 [gr-qc]].
 
\bibitem{DYSON:2013jra}
F.~Dyson,
Int. J. Mod. Phys. A \textbf{28}, 1330041 (2013)

\bibitem{Rothman:2006fp}
T.~Rothman and S.~Boughn,
Found. Phys. \textbf{36}, 1801-1825 (2006)
[arXiv:gr-qc/0601043 [gr-qc]].

\bibitem{Kanno:2017dci}
S.~Kanno and J.~Soda,
Phys. Rev. D \textbf{96}, no.8, 083501 (2017)
[arXiv:1705.06199 [hep-th]].

\bibitem{Kanno:2018cuk}
S.~Kanno and J.~Soda,
Phys. Rev. D \textbf{99}, no.8, 084010 (2019)
[arXiv:1810.07604 [hep-th]].

\bibitem{Kanno:2019gqw}
S.~Kanno,
Phys. Rev. D \textbf{100}, no.12, 123536 (2019)
[arXiv:1905.06800 [hep-th]].

\bibitem{Grishchuk:1974ny}
L.~P.~Grishchuk,
Zh. Eksp. Teor. Fiz. \textbf{67}, 825-838 (1974)

\bibitem{Grishchuk:1990bj}
L.~P.~Grishchuk and Y.~V.~Sidorov,
Phys. Rev. D \textbf{42}, 3413-3421 (1990)
doi:10.1103/PhysRevD.42.3413

\bibitem{Allen:1999xw}
B.~Allen, E.~E.~Flanagan and M.~A.~Papa,
Phys. Rev. D \textbf{61}, 024024 (2000)
[arXiv:gr-qc/9906054 [gr-qc]].

\bibitem{Parikh:2020nrd}
M.~Parikh, F.~Wilczek and G.~Zahariade,
[arXiv:2005.07211 [hep-th]].

\bibitem{Parikh:2020kfh}
M.~Parikh, F.~Wilczek and G.~Zahariade,
[arXiv:2010.08205 [hep-th]].

\bibitem{Parikh:2020fhy}
M.~Parikh, F.~Wilczek and G.~Zahariade,
[arXiv:2010.08208 [hep-th]].

\bibitem{Haba:2020jqs}
Z.~Haba,
Eur. Phys. J. C \textbf{81}, no.1, 40 (2021)
[arXiv:2009.12306 [gr-qc]].

\bibitem{Riedel:2013yca}
C.~J.~Riedel,
[arXiv:1310.6347 [quant-ph]].

\bibitem{Suzuki:2015nva}
F.~Suzuki and F.~Queisser,
J. Phys. Conf. Ser. \textbf{626}, no.1, 012039 (2015)
[arXiv:1502.01386 [gr-qc]].

\bibitem{Kanno:2020usf}
S.~Kanno, J.~Soda and J.~Tokuda,
Phys. Rev. D \textbf{103}, no.4, 044017 (2021)
[arXiv:2007.09838 [hep-th]].

\bibitem{Feynman:1963fq}
R.~P.~Feynman and F.~L.~Vernon, Jr.,
Annals Phys. \textbf{24}, 118-173 (1963)

\bibitem{Breuer}
Heinz-Peter Breuer and Francesco Petruccione
Phys. Rev. A \textbf{63}, 032102

\bibitem{DeLorenci:2014vwa}
V.~A.~De Lorenci and L.~H.~Ford,
Phys. Rev. D \textbf{91}, no.4, 044038 (2015)
[arXiv:1412.4685 [gr-qc]].

\bibitem{Quinones:2017wka}
D.~A.~Quinones, T.~Oniga, B.~T.~H.~Varcoe and C.~H.~T.~Wang,
Phys. Rev. D \textbf{96}, no.4, 044018 (2017)
[arXiv:1702.03905 [gr-qc]].

\bibitem{Oniga:2015lro}
T.~Oniga and C.~H.~T.~Wang,
Phys. Rev. D \textbf{93}, no.4, 044027 (2016)
[arXiv:1511.06678 [quant-ph]].

\bibitem{Oniga:2017pyq}
T.~Oniga and C.~H.~T.~Wang,
Phys. Rev. D \textbf{96}, no.8, 084014 (2017)
[arXiv:1701.04122 [gr-qc]].

\bibitem{Maggiore:1999vm}
M.~Maggiore,
Phys. Rept. \textbf{331}, 283-367 (2000)
[arXiv:gr-qc/9909001 [gr-qc]].

\bibitem{Blencowe:2012mp}
M.~P.~Blencowe,
Phys. Rev. Lett. \textbf{111}, no.2, 021302 (2013)
[arXiv:1211.4751 [quant-ph]].

\bibitem{Anastopoulos:2013zya}
C.~Anastopoulos and B.~L.~Hu,
Class. Quant. Grav. \textbf{30}, 165007 (2013)
[arXiv:1305.5231 [gr-qc]].


\bibitem{Gely:2021fhv}
M.~F.~Gely and G.~A.~Steele,
[arXiv:2103.12729 [quant-ph]].

\bibitem{Yu:2020ece}
H.~Yu \textit{et al.} [LIGO Scientific],
Nature \textbf{583}, no.7814, 43-47 (2020)
[arXiv:2002.01519 [quant-ph]].















    
\end{thebibliography}
\end{document}